\documentclass[a4paper, 11pt]{article}
\date{December 2005}
\usepackage{epsfig}
 
\newcommand{\be}{\begin{equation}}
\newcommand{\ee}{\end{equation}}
\newcommand{\ba}{\begin{eqnarray}}
\newcommand{\ea}{\end{eqnarray}}
\newcommand{\bi}{\begin{itemize}}
\newcommand{\ei}{\end{itemize}}
\newcommand{\tr}{{\rm Tr\,}}

\newcommand{\nn}{\nonumber \\}

\newcommand{\<}{\langle}
\renewcommand{\>}{\rangle}
\newcommand{\eq}{Eq.~}

\newcommand{\la}{\label}

\begin{document}
\begin{titlepage}
\begin{flushright}
DESY preprint 06-005

\end{flushright}
\begin{centering}
\vfill
                                                                                                   
\vspace*{2.0cm}
{\bf \Large Poincar\'e invariance in effective string theories}

\vspace{2.0cm}
{\bf Harvey~B.~Meyer}
\centerline{Deutsches Elektronen Synchrotron DESY}
\centerline{Platanenallee 6}
\centerline{D-15738 Zeuthen}
\vspace{0.1cm}\\
\centerline{harvey.meyer@desy.de}

\vspace*{2.0cm}

\end{centering}
\vfill

\centerline{\bf Abstract}
\vspace{0.1cm}
\noindent 
We investigate the dispersion relation of the winding closed-string states 
in $SU(N)$ gauge theory defined on a $d$-dimensional hypertorus,
in a class of effective string theories.
We show that order by order in the asymptotic expansion, each energy eigenstate satisfies 
a relativistic dispersion relation. This is illustrated 
in the L\"uscher-Weisz effective string theory to two-loop order, 
where the Polyakov loop matrix elements between the vacuum and 
the closed string states are obtained explicitly. 
We attempt a generalization of these considerations to the case of compact
dimensions transverse to the string. 
\vfill
\end{titlepage}

\setcounter{footnote}{0}

\section{Introduction\la{sec:intro}}
The degrees of freedom and the dynamics of the QCD string have been 
the object of detailed studies in recent years (see~\cite{kuti_dublin} for a review).
The energy 
stored in the gauge field in the presence of a distant static $Q\bar Q$ pair grows
linearly with their separation  $R$ (see e.g. \cite{Luscher:2002qv,pushan} for recent data)
in the pure SU($N$) theory, 
suggesting that the flux-lines running between the two color charges behave like
a string with a string tension $\sigma$.

From this hypothesis, it follows in a natural way that 
the leading correction to the linear potential corresponds to Gaussian quantum 
fluctuations of the string around its classical configuration~\cite{luscher81}, 
and leads to a regularly spaced low-energy string spectrum with a gap of 
$\pi/R$. This has been verified numerically with remarkable 
accuracy~\cite{Caselle:1996ii,Juge:2002br,juge-kuti-morning,Caselle:2005xy}.

An elegant way to connect the string-theory predictions with a (gauge-invariant)
observable of the gauge theory is to consider the free energy of the $Q\bar Q$ pair
when inserted inside a thermal heatbath with time extent $T$. 
The latter observable is then the Polyakov 
loop correlator~\cite{Luscher:2002qv,Caselle:2002rm} $ P^*(\mathbf{x}) ~ P(\mathbf{y}) $, 
and a systematic expansion in $(\sigma RT)^{-1}$ 
for its expectation value beyond the area law was proposed in~\cite{lw}. 
The free-string partition function reads~\cite{Caselle:1996ii,lw,Dietz:1982uc}
(in the notation of~\cite{lw}, except for $\tilde E_n \to \tilde M_n$)
\be
{\cal Z}_0(T,r) = e^{-\sigma RT-\mu T} \eta(q)^{2-d},\qquad q\equiv e^{-\pi T/r}
\la{eq:Z0}
\ee
where
\be
\eta(q) = q^{\frac{1}{24}}~\prod_{n=1}^\infty (1-q^n).
\ee
Due to the latter function's modular transformation
property, ${\cal Z}_0$ can be expanded either 
in terms of eigenstates of the open-string or the closed-string sector~\cite{lw},
\ba
{\cal Z}_0(T,r) =\qquad\qquad\qquad\quad& \sum_{n=0}^\infty & w_n ~e^{-E^0_n T} 
                   \la{eq:open_Z}\\
         =  e^{-\mu T}\left(\frac{T}{2r}\right)^{\frac{1}{2}(d-2)}
            & \sum_{n=0}^\infty  & w_n ~e^{-\tilde M^0_n r}.  \la{eq:closed_Z}
\ea
with
\ba
E^0_n(r)         &=& \mu + \sigma r + \frac{\pi}{r}(-\frac{d-2}{24} + n)   \la{eq:open_E}\\
\tilde M^0_n(T)  &=&       \sigma T + \frac{4\pi}{T}(-\frac{d-2}{24} + n)   \la{eq:closed_E}
\ea
the energies of the open and closed string states respectively. 
The low-lying closed-string spectrum has also been investigated 
recently~\cite{Meyer:2004hv,Juge:2003vw}. The next order
of the expansion will be discussed below. 

The idea of a derivative expansion in the transverse fluctuations was 
already present in~\cite{luscher81}. The work of Polchinski and Strominger
\cite{polchinski-strominger} was the first to aim at making predictions beyond the universal $1/r$ 
corrections. It has the merit of being manifestly Poincar\'e 
invariant, and to give the possibility of an insight into how the worldsheet 
degrees of freedom are related to those of the underlying quantum field theory.
The more recent approach of L\"uscher and Weisz~\cite{lw} has the nice feature that once the
two-dimensional field theory has been postulated and the degrees of freedom identified,
the most general interactions, including boundary operators, can be written down automatically.
It seems that the Polchinski-Strominger prescription of plugging the induced metric into the 
Polyakov determinant fixes the coefficient of the $1/r^3$ energy corrections, whereas (in $d=4$)
they are multiplied by an undetermined coefficient in the L\"uscher-Weisz effective theory.

In this paper, we address the question whether the proposed effective theory
is consistent with Poincar\'e invariance. Indeed, string theories such as the Nambu-Goto string
can be formulated in a form where the latter symmetry is manifest. In an
effective string theory the symmetry is in general not manifest. We will however show that
to any finite order in the string expansion all closed string states 
admit a relativistic dispersion relation, ${\rm E}^2(p)= M^2 + p^2$ (this property 
cannot be discussed for the open-string states because they are attached to static charges). 
The proof is simple and does not rely on the details of the effective theory, but only 
on the rotation symmetry of infinite space around the worldline of a static quark, 
and the requirement that open-closed string duality holds order by order in an 
expansion in powers of $1/$distance.

Section 2 describes the setup and gives the proof to the statement made.
Section 3 and 4 contain the explicit calculation of the closed-string 
dispersion relation and Polyakov loop matrix element between the vacuum and the 
energy eigenstates in the L\"uscher-Weisz theory, at leading and at the next order
respectively; the constraints on the theory's unknown coefficients are rederived.
Section 5 generalizes the discussion to the case of compact dimensions transverse
to the Polyakov-loop plane. A conclusion summarizes the results obtained.
\section{Fourier representation of the Polyakov loop correlator}
In this section we focus on the situation 
where all dimensions, except possibly the one around which the Polyakov loops 
wind, are much larger than their separation. For two $d$-dimensional vectors 
\[\mathbf{x}\equiv(0,x_1,\dots,x_{d-2},0) \quad {\rm and} \quad
  \mathbf{y}=(0,0,\dots,0,r),\] 
we have
\be
\< P^*(\mathbf{x}) ~ P(\mathbf{y}) \> = f(\vec x,r,T) \la{eq:f_def}
\ee
where $\vec x=(x_1,\dots,x_{d-2})$ is a $(d-2)$-dimensional vector 
and, in the lattice regularized gauge theory, 
\be
P(x) = \tr\{U_\mu(x)U_\mu(x+a\hat\mu)\dots U_\mu(x+(T-a)\hat\mu )  \}_{\mu=0}.
\ee
\begin{figure*}[t]
\vspace{-0.5cm}
\centerline{\begin{minipage}[c]{10cm}
\psfig{file=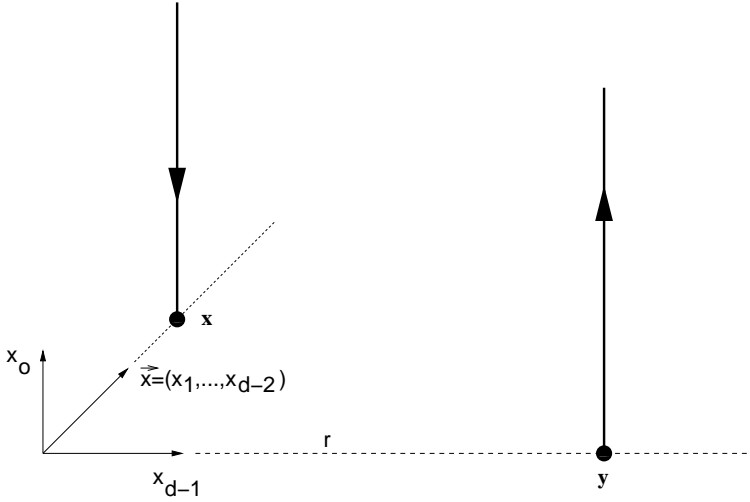,angle=0,width=10cm}       \end{minipage}}
\caption{The setup with two Polyakov loops located at ${\bf x}$ and ${\bf y}$
winding in direction $\mu=0$.}
\la{fig:setup}
\end{figure*}
See figure~\ref{fig:setup}.
We consider the Fourier representation of $f$ with respect to $x$:
\be
 f(\vec x,r,T) = \int \frac{d^{d-2}\vec p}{(2\pi)^{d-2}}~
                          c(\vec p,r,T) ~e^{i\vec p\cdot \vec x},
\la{eq:f_fourier}
\ee
The Fourier coefficients $c$ are conversely given by
\ba
c(\vec p,r,T) &=& \int d^{d-2}\vec x ~
 ~e^{-i\vec p\cdot \vec x} ~ f(\vec x,r,T), \la{eq:f_inv_fourier}\\
   &=&  \frac{1}{V_{d-2}} ~~\< ~{\cal O}^*(0) ~ {\cal O}(r)~  \>.
\ea
where generically
\be
{\cal O}(r) \equiv \int d^{d-2}\vec z ~~e^{i\vec p\cdot \vec z} ~
              P(0,z_1,z_2,\dots,z_{d-2},r)
\ee
and $V_{d-2}$ is the volume of the space parametrized by 
$x_1,\dots,x_{d-2}$.
These coefficients thus represent correlation functions of 
Polyakov loops with a definite momentum $\vec p$ along 
directions 1 through $d-2$.
On general grounds, the 
spectral representation of $c(\vec p,r,T)$ in the gauge theory
(e.g. in terms of eigenstates of the
transfer matrix of the lattice-regularized gauge theory 
along the direction labelled by ${d-1}$) reads
\be
c(\vec p,r,T) = \sum_{n\geq0} |v_n(\vec p,T)|^2~e^{-\tilde {\rm E}_n(\vec p,T)r}.
\la{eq:transf_mat}
\ee
Thus the dispersion relation of the winding flux-tube states
can be read off from this quantity.

The $SO(d-1)$ rotational symmetry around one 
Polyakov loop implies that the correlator is a function of a 
reduced number of variables:
\be
f(\vec x,r,T) = \varphi(\sqrt{r^2+\rho^2},T),
\qquad \rho\equiv |\vec x|.
\la{eq:varphi}
\ee
Due to the $SO(d-2)$ symmetry in the dimensions $1,\dots,d-1$,
$c$ obviously depends on $\vec p$ only through $|\vec p|$, 
and so do $v_n$ and ${\rm E}_n$. How the remaining 
$SO(d-1)/SO(d-2)\simeq SO(2)$ symmetry is reflected in 
these functions is the subject of the following section.

\subsection{Boost invariance and the two-point function of a local operator\la{sec:boost}}
In this section, we drop the $T$-dependence of $c$, $v_n$ and ${\rm E}_n$,
because it will play no role. In particular, what follows applies
just as well to the two-point function of QCD operators such as
$\tr F_{\mu\nu}^2$, $\bar\psi \Gamma \psi$, etc. in $d-1$ dimensions.
Starting from~\eq\ref{eq:f_inv_fourier} with $f$ given by~(\ref{eq:varphi}), 
it is easy to show that 
\be
\frac{\partial}{\partial r}\frac{\partial}{\partial p}~ c(p,r) = pr~ c(p,r)
\la{eq:condition1}
\ee
Conversely, if $c(\vec p,r)$ satisfies \eq\ref{eq:condition1}, 
it is clear that the function $f$ defined by \eq\ref{eq:f_fourier} satisfies
$ (\rho \partial_r - r \partial_\rho) f(\rho,r)=0$, 
which implies that $f$ is of the form~(\ref{eq:varphi})\footnote{Indeed, $f$ is then
invariant under an infinitesimal rotation $(\rho,r)\to  (\rho+\theta r,r-\theta \rho).$}. 
Thus no information
about the symmetry has been lost in going from (\ref{eq:varphi}) 
to (\ref{eq:condition1}).

For a function $c$ admitting an absolutely convergent
spectral representation as in (\ref{eq:transf_mat}),
\eq\ref{eq:condition1} is equivalent to
\be
\sum_{n\geq0} e^{-\tilde {\rm E}_n(p) r} \left[
|v_n(p)|^2\left(1-\frac{d}{d(p^2)} \tilde {\rm E}_n^2(p)\right)
+\frac{2}{r} \frac{d}{d(p^2)}\left( |v_n(p)|^2 \tilde {\rm E}_n(p)\right)\right] = 0,
\quad\forall (p,r).
\la{eq:condition2}
\ee
In particular,
dividing by $e^{-\tilde {\rm E}_0(p)r}$, and letting $r\to\infty$, 
only the  $n=0$ term remains. 
In this way we learn that, for $n=0$,  
\be
\frac{d}{d(p^2)} \tilde {\rm E}_n^2(p) = 1,\quad {\rm i.e.}\quad  
\tilde {\rm E}_n^2(p) = \tilde {\rm E}_n^2(0) + p^2,  \la{eq:e_rel}
\ee
and the momentum dependence of the matrix elements $v_0$ is also fixed:
\be
|v_n(p)|^2  = \frac{ b_n } { \tilde {\rm E}_n(p) }.  \la{eq:v_rel}
\ee
Thus the $n=0$ term does not contribute to \eq\ref{eq:condition2}.
One may then iterate the argument for $n=1,2,\dots$, 
showing that \eq\ref{eq:e_rel} and \ref{eq:v_rel} hold for all states.
If say a double degeneracy occurs, one can  only conclude that the sum of the 
$|v_n(p)|^2$ are inversely proportional to the common energy. However exact degeneracies
normally only occur in quantum field theory for symmetry reasons, in which case
one also expects the $|v_n(p)|^2$ of the two states to be equal.

One may draw two conclusions:
the states that contribute to a correlator such as  $c$ defined in \eq\ref{eq:f_inv_fourier}
with $f(\vec x,r)$ of the form (\ref{eq:varphi})
necessarily admit a relativistic dispersion relation.
Conversely, for relativistic $|v_n(p)|^2$ and $\tilde {\rm E}_n(p)$,
$f(\vec x,r)$ is of the form (\ref{eq:varphi}); 
which implies in particular a central potential for the $Q\bar Q$ pair.

It is instructive to ask oneself which step of the argument 
fails to hold in the case of a non-relativistic theory. In that case, 
the Euclidean time-direction plays a special role and cannot be interchanged
for a spatial coordinate. Thus if $x_0$ is taken to be the time-direction, 
then \eq\ref{eq:varphi} holds by spatial rotation symmetry,
but \eq\ref{eq:transf_mat} does not hold in general, and even if it does
the $\tilde E_n(p)$ are not the eigenvalues of the Hamiltonian, 
which governs the evolution of states in the $x_0$ direction rather than in
the $x_{d-1}$ direction. Alternatively, if it is $x_{d-1}$ that 
plays the role of time, then \eq\ref{eq:varphi} does not hold
($\rho$ is now a spatial and $r$ a temporal separation), 
precisely because the theory does not have the relativistic boost invariance.

\subsection{Boost invariance in an effective string theory\la{sec:ld}}
As we shall see in the explicit calculations of sections 3 and 4, 
the L\"uscher-Weisz effective string theory, worked out to order $m$, 
generically produces an expression for $c(p,r)$ of the form
\be
c_m(p,r) = \sum_n e^{-\tilde{\rm E}_n(p)r}\sum_{k=-k_m}^\infty
\frac{\gamma^{(m)}_{nk}(p)}{r^k}.
\ee
Here we assume that this correlator has been obtained 
by performing an integral of the type (\ref{eq:f_inv_fourier}) 
with $f(\vec x,r)$ of the form (\ref{eq:varphi}).
Then \eq\ref{eq:condition1} implies $\beta_{nk}(p)=0$ $\forall n,k$ where
\ba
\beta_{nk}(p) &\equiv& \gamma^{(m)}_{nk}(p) 
\left(1-\frac{d}{d(p^2)}   \tilde {\rm E}^2_n(p)\right)
  + \frac{2}{\tilde{\rm E}_n(p)^{1-k}}\frac{d}{d(p^2)}
 \left(\tilde{\rm E}_n(p)^{2-k}\gamma^{(m)}_{n,k-1}(p)\right)  \nn
&&   + 2(k-2) \frac{d}{d(p^2)} \gamma^{(m)}_{nk-2}(p), \qquad k\geq -k_m\la{eq:beta_nk}
\ea
with the understanding $\gamma^{(m)}_{n,-k_m-1}=\gamma^{(m)}_{n,-k_m-2}=0$.
For $k=-k_m$, the equation teaches us that 
$\tilde{\rm E}_n(p)^2 = \tilde M_n^2+p^2$.

It is a straightforward exercise to show by induction that these conditions imply
\be
\gamma^{(m)}_{nk} (p) = \tilde{\rm E}_n(p)^{k-1} \sum_{j=0}^{k+k_m}
\frac{\alpha^{(m)}_{n,k-j} (k-1)(k-2)\dots (k-2j)}{j!(2\tilde{\rm E}^2_n(p))^j}
\ee
where the $\alpha^{(m)}_{nk}$ ($k\geq -k_m $) are real numbers.
\subsubsection{Energy corrections}
The computational scheme is meant to hold order by order in powers of inverse distance, 
so that the terms with positive powers of $r$ in $c(p,r)$ must be interpreted 
as giving the energy shifts of the closed-string states. 
It is not \emph{a priori} obvious that the argument of section~\ref{sec:boost}
goes through, since the terms of the exponential that are missing are not uniformly small in
$r$. We will nevertheless show that the energy shifts preserve the relativistic 
dispersion relation.

If the scheme is to be at all consistent, 
the energy shifts must be given by the ratio of the $r^1$ to the $r^0$ coefficients:
\be
\delta \tilde {\rm E}^{(m)}_n(p) =
 -\frac{\gamma^{(m)}_{n,-1}(p)}{ \gamma^{(m)}_{n,0}(p)  }  \la{eq:dEgg}
\ee
The understanding is that the fraction should be expanded in powers of $\tilde {\rm E}_n(p)^{-1}$
and that the terms beyond the $k_m^{\rm th}$ power ought to be neglected.
For the quadratic, cubic, etc. terms to be consistent with the Taylor expansion 
of the exponential function, one finds that the numerical coefficients $\alpha^{(m)}_{nk}$
must be related by
\be
\frac{\alpha^{(m)}_{n,-k}}{\alpha^{(m)}_{n0}}= 
\frac{1}{k!}\left(\frac{\alpha^{(m)}_{n,-1}}{\alpha^{(m)}_{n0}}\right)^k, \qquad k\geq 0.
\ee
Thus 
\be
\gamma^{(m)}_{n,-k}(p)=\frac{\alpha^{(m)}_{n0}}{\tilde {\rm E}_n(p)^{k+1}k!}
\left(\frac{\alpha^{(m)}_{n,-1}}{\alpha^{(m)}_{n0}}\right)^k 
\sum_{j=0}^{k_m-k}  \frac{(2j+k)!}{j!(j+k)!}
\left(\frac{\alpha^{(m)}_{n,-1}}{2\alpha^{(m)}_{n0}\tilde {\rm E}_n^2(p)}\right)^j
\quad 0\leq k\leq k_m. \la{eq:gg}
\ee

On the other hand, if $E(p)$ satisfies the relativistic dispersion relation,
then so does $E(p)(1+x(p))$ if and only if 
\be
x(p) = -1 + \sqrt{1+x_0(x_0+2) (M/E(p))^2},
\ee
where $x_0\equiv x(0)$ and $ M = E(p=0)$.
It is now a lengthy but straightforward exercise 
to check that \eq\ref{eq:dEgg} and \ref{eq:gg} satisfy this condition
up to order $1/\tilde{\rm E}_n^2(p)^{k_m+1}$ corrections and as a by-product one finds 
\be
-\frac{\delta \tilde M_n^{(m)}}{\tilde M_n} = \sum_{j=1}^{k_m} 
\left( \frac{\alpha^{(m)}_{n,-1}}{\alpha^{(m)}_{n0}\tilde M_n^2}\right)^j \frac{(2j-3)!!}{j!}.
\la{eq:M_n}
\ee
(with the convention $(-1)!! = 1$). At this stage, one may consider 
that the series is actually of the form (\ref{eq:c_m}), where the energy levels
must now be understood as the ones including the corrections (\ref{eq:M_n}).
The analysis of the next section thus applies.

In summary, the type of expressions one obtains for $c(p,r)$
from a long-distance effective theory for a sector of very massive states
is such that the latter automatically admit a relativistic dispersion relation
as soon as the expression is made consistent with the form expected from 
a spectral representation. In general these consistency requirements will 
constrain the parameters of the effective theory. We saw that rotation 
invariance implies that the $\gamma^{(m)}_{n,-k}$, $0\leq k\leq k_m$
are parametrized in terms of $(k_m+1)$ numbers $\alpha^{(m)}_{n,-k}$, $0\leq k\leq k_m$.
The requirement of the exponentiation of these terms leads to $k_m-1$
conditions, so that to any order, there are exactly two free parameters per energy 
level, $\alpha^{(m)}_{n0}$ and $\alpha^{(m)}_{n,-1}$, which determine the 
overlap and energy of each state $n$. 
If the effective theory can satisfy these  $k_m-1$ conditions,
the  scheme is thus consistent to all orders.

\subsubsection{Cancelling powers of $1/r$}
We now suppose that the string-theory parameters have been tuned so as to give
a definite-momentum correlator of the form
\be
c_m(p,r) = \sum_n e^{-\tilde {\rm E}^{(m)}_n(p) r } \sum_{k=0}^\infty
\frac{  \gamma^{(m)}_{nk}(p)}{r^k}. \la{eq:c_m}
\ee
with the $\tilde{\rm E}^{(m)}_n(p)$ satisfying the relativistic dispersion relation.
The functional form is only consistent with the 
spectral representation (\ref{eq:transf_mat}) if one is able to further tune
the free parameters of the theory to cancel the terms
$k=1,\dots,k_m$, and one then neglects the terms with $k>k_m$. 
This tuning is possible for the first two orders
of the L\"uscher-Weisz theory (see \cite{lw} and the next two sections).
\eq\ref{eq:condition1} applied to the form (\ref{eq:c_m}) leads
to \eq\ref{eq:beta_nk} with $k_m=0$.
As we assume $\gamma^{(m)}_{n0}(p)\neq0$, $\beta_{n0}(p)=0$ requires the relativistic
dispersion relation (\ref{eq:e_rel}), and the next equation
then implies (\ref{eq:v_rel}). At $k=1$, one finds
that $\gamma^{(m)}_{n1}(p)$ is independent of $p$; note that 
the boost symmetry does not imply a relation between the $\alpha_{n,k>0}$
and $\alpha_{n,k\leq0}$.
Further on, it is clear that if $\gamma^{(m)}_{nk}(p)=0$ for $k=1,\dots,m-1$, 
then the momentum dependence of the next coefficient is very simple:
\be
\gamma^{(m)}_{nk}(p) = \gamma^{(m)}_{nk}(0) 
~\left( \frac{\tilde {\rm E}^{(m)}_n(p)}{\tilde M_n}  \right)^{k-1}.
\ee
In particular, it vanishes identically if and only if it vanishes at $p=0$.

The conclusion is that for any effective theory of $f(\vec x, r, T)$ 
which produces an expression of the type (\ref{eq:c_m}) with
$\gamma^{(m)}_{n0}(p)\neq0$, rotation symmetry automatically implies 
the  relations $\tilde {\rm E}^{m}_n(p) = \sqrt{p^2 + \tilde (M^m_n)^2}$
and $\gamma^{(m)}_{n0}(p) \propto (\tilde {\rm E}^m_n(p))^{-1}$ and also determines
the momentum dependence of $\gamma^{(m)}_{nk\geq1}(p)$.
Cancelling systematically the latter in the effective theory 
thus allows one to interpret $\tilde {\rm E}_n(p)$ as the energies of relativistic closed strings.
This can be done order by order consistently with Euclidean rotation symmetry,
the technical reason being that the derivative w.r.t. and the multiplication by $r$ in 
the identity (\ref{eq:condition1}) act locally on the power series in $1/r$.

\section{The free bosonic string theory}
The effective theory~(\ref{eq:Z0}) makes a prediction for the Polyakov loop correlator
(\ref{eq:f_def}) for any separation. In this section 
we exploit this fact to compute the Polyakov
correlators of definite momentum explicitly, 
thus extracting the predicted dispersion relation for the closed string states.
\paragraph{4 dimensions\\}
Since the spatial dimensions have $SO(3)$ as symmetry group, 
$c_0(\vec p,r,T)$, the  free-string theory
prediction for  $c(\vec p,r,T)$, can be computed straightforwardly:
\be
c_0(\vec p,r,T) = e^{-\mu T}~ \int d^2\vec x ~
 e^{-i\vec p\cdot \vec x} \frac{T}{2\sqrt{|\vec x|^2 + r^2}}~\sum_{n\geq0} w_n~
e^{-\tilde M^0_n\sqrt{r^2+|\vec x|^2}}.
\ee
We obtain
\be
c_0(p,T,r) = e^{-\mu T} ~\pi T ~\sum_{n\geq0} w_n~\frac{1}{\sqrt{p^2+(\tilde {M^0_n})^2}}
       e^{-r\sqrt{p^2+(\tilde {M^0_n})^2}}, \la{eq:c_4d_exact}
\ee
from which we read off
\ba
d=4:\qquad \tilde {\rm E}^0_n(p,T) & = & \sqrt{p^2+(\tilde {M^0_n})^2} \\
|v_n(p,T)|^2 & = &e^{-\mu T}~\frac{\pi T w_n}{ \sqrt{p^2+(\tilde {M^0_n})^2}}.
\ea
As expected, the effective string theory predicts exactly a free, relativistic dispersion
relation for the closed string states. The situation is special in four dimensions
in that there are no inverse powers of $r$ appear in $c(p,r,T)$ to this order.
\paragraph{3 dimensions\\}
Repeating the exercise for $d=3$,
\be
c_0(p,r,T) = e^{-\mu T} ~\int_{-\infty}^\infty dx~ e^{-ipx}
\left(\frac{T}{2\sqrt{r^2+x^2}}\right)^{1/2}
 \sum_{n\geq0} w_n e^{-\tilde M^0_n\sqrt{r^2+x^2}},
\ee
we find 
\be
c_0(p,r,T) = e^{-\mu T} ~r ~\sqrt{\frac{T}{4\pi}} ~\sum_{n\geq0} 
w_n~ \frac{(\tilde {M^0_n})^{3/2}}{\sqrt{p^2+(\tilde {M^0_n})^2}} \\
\left( K_{\frac{3}{4}}(z_n^+)K_{\frac{1}{4}}(z_n^-) 
+     K_{\frac{3}{4}}(z_n^-)K_{\frac{1}{4}}(z_n^+)\right)
\la{eq:c_3d_exact}
\ee
where
\be
z_n^\pm \equiv \frac{r}{2}\left( \sqrt{(\tilde {M^0_n})^2 + p^2} \pm |p|\right). 
\ee
At this stage one can verify that expression~(\ref{eq:c_3d_exact})
satisfies~(\ref{eq:condition1}).
Using the asymptotic expansion for the modified Bessel functions yields 
\be
c_0(p,r,T)\sim e^{-\mu T} ~\sqrt{\pi T} ~\sum_{n\geq0} ~w_n~ 
\sqrt{\frac{\tilde M^0_n}{(\tilde {M^0_n})^2+p^2}} ~e^{-r\sqrt{(\tilde {M^0_n})^2+p^2}}
\left(1 + \frac{\sqrt{(\tilde {M^0_n})^2+p^2}}{8 r (\tilde M_n^0)^2}+\dots\right).
\la{eq:c_3d_asympt}
\ee
Keeping only the leading order term, we have:
\ba
d=3:\qquad  \tilde {\rm E}^0_n(p,T) & = & \sqrt{p^2+(\tilde {M^0_n})^2} \\
    |v_n(p,T)|^2 &= & e^{-\mu T} w_n~\sqrt{\frac{\pi T \tilde M^0_n}{p^2+(\tilde M^0_n)^2}}.
\ea
Thus the leading asymptotic form of $c_0$ still satisfies \eq\ref{eq:condition2}. 

\section{The L\"uscher-Weisz theory at two-loop order}
In this section we shall compute the leading non-trivial corrections
to the spectrum and matrix elements in the closed string sector
predicted by the L\"uscher-Weisz effective theory.
Ref.~\cite{lw} gives the general expression for the partition function at two-loop order:
\be
{\cal Z}(r,T) = \sum_{n\geq 0} w_n \{1-[E_n^1]T \} e^{-E_n^0 T},
\ee
with $[E_{n}^1]$ the weighted average of the first-order energy shifts $E_{n,i}^1$
within the $n^{\rm th}$ free-string multiplet.
\paragraph{4 dimensions\\}
For general values of the `low-energy constants' (called $c_2$ and $c_3$), 
the corresponding closed-string expansion reads~\cite{lw}
\be
{\cal Z}(r,T) = e^{-\mu T}\frac{T}{2r} 
               \sum_{n\geq0} w_n~ e^{-\tilde M_n^0 r}
\left\{ 1-[\tilde M_n^1]r +(2c_2+c_3)\left[\frac{4\pi}{T^2}(n-{1\over12})+\frac{1}{2Tr}\right]\right\}.
\ee
Starting from this expression, 
we can compute as before the expression of $c_1(p,r,T)$, the Fourier coefficient of
${\cal Z}(r,T)$ with respect to the transverse coordinates:
\ba
&&c_1(p,r,T) =  e^{-\mu T}~\pi rT~\sum_{n\geq0}w_n ~\times \\
&& \left\{ 1 + r[\tilde M_{n}^1] \frac{\partial}{\partial u}
+(2c_2+c_3)\frac{4\pi}{T^2}(n-{1\over12})+\frac{2c_2+c_3}{2Tr}\int_{r\tilde M_n^0}^\infty du
 \right\}_{u=r\tilde M_n^0}^{v=pr}   I^{(4)}(u,v)  \nonumber
\ea
with
\be
 I^{(4)}(u,v) = \frac{e^{-\sqrt{u^2+v^2}}}{\sqrt{u^2+v^2}}.
\ee
What powers of $r$ appear in this expression? To answer this question one may first
inspect the $v=0$ case (zero momentum).
It is easy to see that, while the free theory term gives a contribution $O(r^0)$, 
the second term makes contributions $O(r^0)$ and $O(r^{+1})$, and the fourth one produces
a whole series of inverse powers of $r$ starting at $r^{-1}$. While positive powers of $r$,
once exponentiated, give us the energy correction, the $1/r$ term is inconsistent 
with the spectral representation of $c$. Therefore, requiring that such terms be absent
yields the constraint $2c_2+c_3=0$ (already obtained in~\cite{lw}). 
It is remarkable that there are then no inverse powers of $r$ left at all.

For this special set of parameters, the closed-string energy corrections are~\cite{lw}
\be
d=4: \qquad  \tilde M_{n}^1(T) =  \tilde M_{n}^0 + [\tilde M_{n}^1],\qquad
[\tilde M_{n}^1](T) = -c_2\frac{(4\pi)^2}{T^3}(n-\frac{1}{12})^2.
\ee
The final result  is 
\be
c_1(p,r,T) =  \sum_{n\geq0} ~ |v_n^1(p,T)|^2 ~
e^{-r \tilde {\rm E}_n^1(p,T)} ~+~ O([\tilde M_{n}^1]^2)
\ee
where 
\ba
\tilde {\rm E}_n^1(p,T) &=& \sqrt{p^2 +(\tilde M_{n}^0)^2}
  + \frac{\tilde M_n^0 ~[\tilde M_{n}^1]}{\sqrt{p^2 +(\tilde M_n^0)^2}}  
\la{eq:tE_1}\\
|v_n^1(p,T)|^2 &=& |v_n^0(p,T)|^2 ~
  \left(1-\frac{\tilde M_n^0  ~ [\tilde M_{n}^1]}{p^2 + (\tilde M_n^0)^2}\right).
\la{eq:v_n_1}
\ea
\eq\ref{eq:tE_1} implies that the relativistic dispersion relation is still 
satisfied to this order:
\be
(\tilde {\rm E}_n^1(p,T))^2 =  p^2 + (\tilde M_{n}^1)^2 ~+~ O([\tilde M_{n}^1]^2),~~\forall n.
\ee
Furthermore, we have
\be
|v_n^1(p,T)|^2 ~\tilde {\rm E}_n^1(p,T) = |v_n^0(p,T)|^2 ~\tilde {\rm E}_n^0(p,T) 
 ~\left[1+ O\left(\left(\frac{[\tilde M_{n}^1]}{\tilde M_{n}^0}\right)^2\right)\right],~~\forall n.   \la{eq:v_times_e}
\ee
Thus $\frac{d}{dp^2}(|v_n^1|^2~\tilde {\rm E}_n^1)$ vanishes to linear order in the energy correction,
as expected from the arguments of section~\ref{sec:ld}.
The corrections to the matrix elements is $O(1/(\sigma T^2)^2)$.
\paragraph{3 dimensions\\}
In 3 dimensions, the closed-string expansion reads~\cite{lw}
\be
{\cal Z}(r,T) = e^{-\mu T}\sqrt{\frac{T}{2r}}\sum_{n\geq 0} w_n e^{-\tilde M_n^0 r} 
 \left\{1-[\tilde M_n^1]r  - c_2  \left[ \frac{4\pi}{T^2}(n-{1\over24})
 +  \frac{1}{4Tr}  \right] \right\} . \la{eq:Z_3d_2loop_closed}
\ee
The calculation is slightly more involved than in 4d. One can first
write the Fourier coefficient in the form:
\ba
c_1(p,r,T) &=& e^{-\mu T} ~\sqrt{\frac{Tr}{4\pi}}~\sum_{n\geq0} ~w_n~\times \la{eq:c_1_3d}\\
&&   \left\{   1-\frac{4\pi c_2}{T^2}(n-{1\over24})
+ r [E_n^1] \frac{\partial}{\partial u} -\frac{c_2}{4Tr}\int_{r\tilde M_n^0}^\infty du
 \right\}_{u=r\tilde M_n^0}^{v=pr}  I^{(3)}(u,v)  \nonumber
\ea
where
\be
I^{(3)}(u,v) \equiv \frac{u^{3/2}}{\sqrt{u^2+v^2}}~
\left( K_{\frac{3}{4}}(z_+)K_{\frac{1}{4}}(z_-)
+     K_{\frac{3}{4}}(z_-)K_{\frac{1}{4}}(z_+)\right),
\ee
$z_\pm$ being the solutions of the quadratic equation
\be
z^2 ~-~ z\sqrt{u^2 + v^2} ~+~ \frac{u^2}{4}  ~=~ 0.
\ee
The primitive of $I^{(3)}(u,v)$ is known: $\frac{\partial}{\partial u}J^{(3)}(u,v)=I^{(3)}(u,v)$ with
\be
J^{(3)}(u,v) = -2\sqrt{u}~K_{1\over4}(z_+) K_{1\over4}(z_-).
\ee
The result can thus be expressed concisely as
\ba
c_1(p,r,T) &=& e^{-\mu T} ~\sqrt{\frac{Tr}{4\pi}}~\sum_{n\geq0} ~w_n~\times \\
&&   \left\{
\left[1-\frac{4\pi c_2}{T^2}(n-{1\over24})\right]\frac{\partial}{\partial u}
+ r [E_n^1] \frac{\partial^2}{\partial u^2}  +\frac{c_2}{4Tr} \right\}_{u=r\tilde M_n^0}^{v=pr}
 J^{(3)}(u,v)  \nonumber
\ea
It is now most efficient to first expand $J^{(3)}(u,v)$ before applying the various 
derivatives. Requiring that the leading $1/r$ term (which comes from the first 
and the last term in the curly brackets) vanishes leads to the 
condition $c_2=\frac{1}{2\sigma}$, thus confirming the result obtained by L\"uscher
and Weisz~\cite{lw}. Some further algebra then leads again to 
the result~(\ref{eq:tE_1}) for the dispersion relation with the energy corrections
given by
\be
[\tilde M_n^1] = -\frac{(4\pi)^2}{2\sigma T^3}(n-{1\over24})^2,
\ee
while the  correction to the matrix elements is 
\ba
&&|v_n^1(p,T)|^2 = |v_n^0|^2~\left(1-\frac{2\pi}{\sigma T^2}(n-{1\over24}) +
\frac{[\tilde M_n^1]}{8 \tilde M_n^0} \frac{3p^2-5(\tilde M_n^0)^2}{p^2+(\tilde M_n^0)^2} \right)\\
& &= |v_n^0|^2~\left(\left(1-\frac{\pi}{\sigma T^2}(n-{1\over24})\right)^2 + 
\left(\frac{2\pi}{\sigma T^2}(n-{1\over24})\right)^2~\frac{(\sigma T)^2 - p^2 }{(\sigma T)^2 + p^2}
~+~O(1/(\sigma T^2)^3) \right). \nonumber
\ea
It is straightforward to verify that 
$\frac{d}{dp^2}(|v_n^1|^2~\tilde {\rm E}_n^1)$ is again $O([\tilde M_n^1]^2)$.
%
\paragraph{}
In the light of the examples seen, it is clear that the general prescription
to obtain the dispersion relation from the closed-string representation of the 
partition function can be given: a positive power $m$ of $r$ inside the curly braces of
\eq\ref{eq:Z_3d_2loop_closed} is replaced 
by $(-r\frac{\partial}{\partial u})^m$ inside the curly braces of \eq\ref{eq:c_1_3d}
and a negative power $m$ of $r$ is replaced by
$\frac{r^{-m}}{m!}\int_{r\tilde M_n^0}^\infty du_1\dots du_m$. 
In general, both lead to energy corrections and to terms containing a negative power
of $r$. The latter must be required to cancel order by order.

\section{The free string with one finite transverse dimension}
In this section, we consider the possibility of having one dimension 
(say $\hat 1$)
transverse to the plane defined by the two Polyakov loops finite and of 
length $L$. The parameter $L^{-1}$ may then also be interpreted as a temperature
at which the gauge theory is probed. In that respect, we remark that 
studying the dynamics of the `spatial' QCD string has 
the advantage (over the ordinary electric one) that the closed strings can 
be kept arbitrary long independently of $L$, while winding modes around 
that periodic dimension play an important role, as we shall see.

In~\cite{hm}, it was argued, using the XY model, 
that when $L$ is finite but large compared to
$\sigma^{-1/2}$, the massless fluctuations of the string are unaffected 
by the size of $L$: the corrections would appear only in 
terms in $E_n$ suppressed by a power of $R$ greater than 3.
The string tension itself, however, becomes a function of $L$.
Thus it is legitimate in this regime to investigate the properties of the `spatial'
Polyakov loop correlator using the free-string partition function.

It is then natural to ask what one obtains for ${\cal Z}_0$ if, working 
backwards, one uses the expression~(\ref{eq:c_4d_exact})
for $c_0$ (which was derived from ${\cal Z}_0$ when $L=\infty$) and sums over
the now discretized momenta $p_1=0,\pm\frac{2\pi}{L},\dots$ 
(compare with \eq\ref{eq:f_fourier}):
\ba
{\cal Z}_0(x_1,r,T,L) &=& e^{-\mu T} \pi T \sum_{n\geq0} w_n \frac{1}{L}\sum_{p_1}
~ e^{ip_1 x_1}~\int_{-\infty}^{\infty} \frac{dp_2}{2\pi}
\frac{e^{-r\sqrt{p_1^2+p_2^2+(\tilde {M^0_n})^2}}}{\sqrt{p_1+p_2^2+(\tilde {M^0_n})^2}} .
\ea
Upon integrating over $p_2$ and using
Poisson's summation formula 
\[ \frac{1}{L}\sum_{p} f(p)=\sum_{m\in \mathbf{Z}}
\int_{-\infty}^{\infty} \frac{dp}{2\pi} e^{impL} f(p),\]
one then finds 
\be
{\cal Z}_0(x_1,r,T,L) = \sum_{m\in \mathbf{Z}} {\cal Z}_0(\sqrt{(mL+x_1)^2+r^2},T).
\la{eq:Z_winding}
\ee
To interpret this result, let us choose $x_1=0$ and 
go to the open string representation of ${\cal Z}_0(r,T)$.
The  energy eigenstates are
\be
E_{mn}(r,L) = \mu + \sigma(L)\sqrt{r^2+m^2L^2} + \frac{\pi}{\sqrt{r^2+m^2L^2}}
              \left(-\frac{1}{24}(d-2)+n\right),\quad n\geq0,m\in \mathbf{Z}
\ee
and they have weight $w_n$ for all $m$.  In particular, the  partition function
(\eq\ref{eq:Z_winding}) is consistent with open-closed string duality.
Given that the direction $L$ was taken 
as periodic, it is clear that we have obtained a sum over classical configurations
(labelled by $m$) of the open string joining the two static charges along 
straight paths winding an arbitrary number of times $m$ around the periodic
dimension, with vibrational states (labelled by $n$) 
built up on each of these classical string configurations.
The same result~(\ref{eq:Z_winding}) is obtained in three dimensions (starting from the 
asymptotic form~\ref{eq:c_3d_asympt}).
The dual nature of the open and closed string expansions of ${\cal Z}_0$
is manifest here, for having \emph{fewer} momenta in the closed string channel
corresponds to \emph{more} states in the open string channel.

We have however not taken into account the fact that the pure SU($N$) gauge theory
has a centre symmetry which is unbroken for $L\gg \sigma^{-1/2}$
and thus forbids states of different ${\cal N}$-ality
to mix. Since the `spatial' Polyakov loops are invariant under this symmetry,
the open-string states can only have winding numbers that are multiples of $N$.
On the closed-string side, this would however imply that the quantum of momentum
has to be $\frac{2\pi}{NL}$ rather than $\frac{2\pi}{L}$. 
It is presently unclear to the author if the states with a fraction
of the normal unit of momentum can be given a sensible interpretation,
for instance as winding modes of the closed strings\footnote{When $L<1/T_c$, 
where $T_c$ is the deconfining temperature
of the gauge theory, the centre symmetry is broken, so this difficulty is absent.}.

Finally, we remark
that the winding modes do not generate a phase transition on the worldsheet.
The reason is that their entropy factor is too weak. Rather we expect worldsheet-tearing
configurations such as the vortices of the XY model to drive the phase transition~\cite{hm}
where the central charge of the string theory drops from $\frac{\pi}{12}$ 
to $\frac{\pi}{24}$ in $d=4$ (or from $\frac{\pi}{24}$ to zero  in $d=3$).
Provided the dimensionally reduced action for hot QCD
correctly describes the asymptotic high-temperature
behaviour of its magnetic observables, then this phenomenon must occur
at a certain $L^*$~\cite{hm}. This is currently being tested numerically~\cite{hm_pushan}.

\section{Conclusion}
We have shown explicitly that the effective string theory of L\"uscher and Weisz
predicts a relativistic dispersion relation for the closed winding flux states
of SU($N$) gauge theories in 3 and 4 dimensions up to two-loop order included.
The matrix elements of Polyakov loop
operators between them and the vacuum were derived as a by-product.
Assuming the closed-string spectrum to be unchanged in the presence of a
compact transverse dimension directly led to the existence of winding modes 
of the open strings around the compact dimension.

At a more general level,
we have considered a long-distance effective theory for the Polyakov loop correlator
in SU($N$) gauge theory  that imposes consistency with the spectral representation
in the crossed channel (closed-string representation)  order by order in the inverse 
Euclidean separation. We have shown that this is sufficient to insure 
relativistic kinematics for the energy eigenstates of that channel.
It is well-known that bosonic string theory, as a theory valid
for all distance scales, is only consistent with Lorentz symmetry in 26 dimensions.
Nevertheless, in a non-critical number of dimensions, 
bosonic string theory can be reconciled  with space-time symmetry
order by order in an expansion around classical, elongated string configurations,
a conclusion already reached in~\cite{polchinski-strominger}.

\paragraph{}
I am very grateful to Rainer Sommer for a discussion which led to a 
           rewriting of section 2.

\end{document}